\documentclass[aps,prl,reprint,superscriptaddress,nobalancelastpage]{revtex4-2}

\usepackage[T1]{fontenc}
\usepackage[utf8]{inputenc}

\usepackage{graphicx}
\usepackage{amsmath,amssymb}
\usepackage{bm}
\usepackage{hyperref}

\begin{document}

\title{Real-Time Formation of a Landau Polaron}

\author{Priya Nagpal}
\affiliation{Department of Chemistry, McGill University, Montreal, H3A 0B8, Canada}

\author{Arnab Ghosh}
\affiliation{Department of Chemistry, McGill University, Montreal, H3A 0B8, Canada}

\author{H\'el\`ene Seiler}
\affiliation{Freie Universit\"at Berlin, Arnimallee 14, 14195 Berlin, Germany}

\author{Samuel Palato}
\affiliation{Department of Chemistry, Humboldt University, Berlin 12489, Germany}

\author{Patanjali Kambhampati}
\affiliation{Department of Chemistry, McGill University, Montreal, H3A 0B8, Canada}

\begin{abstract}
Polarons—electronic excitations dressed by a self-consistent lattice distortion—are fundamental quasiparticles in condensed-matter systems, yet their formation has not been resolved in real time. We develop a microscopic lineshape framework that connects the growth of a collective lattice polarization to the population-time evolution of the anti-diagonal linewidth in coherent multidimensional spectroscopy. Within this formalism, the anti-diagonal linewidth directly tracks the decay of lattice frequency–frequency correlations. Underdamped phonon modes yield oscillatory linewidth modulation, whereas overdamped collective polarization dynamics produce monotonic exponential broadening. Applying this criterion to multidimensional measurements on perovskite quantum dots reported in 2019, we show that the observed ~150 fs exponential anti-diagonal broadening reflects the decay of a collective polarization order parameter. These results establish anti-diagonal linewidth dynamics as a direct real-time signature of Landau polaron formation.
\end{abstract}

\maketitle

Polarons — electronic excitations dressed by a self-consistent lattice distortion\textsuperscript{1-4} — are fundamental quasiparticles in condensed-matter systems\textsuperscript{3,5-9}. Their theoretical description, originating with Landau \textsuperscript{1} and developed through subsequent treatments of electron–phonon coupling \textsuperscript{10-12}, establishes that quasiparticle formation arises from the dynamical reorganization of the surrounding lattice. Yet experimental evidence has largely been limited to steady-state or long-time observables. Transport renormalization, dielectric response, and static spectral shifts reveal the presence of a dressed excitation \textsuperscript{13-17} but do not resolve the process by which the lattice distortion forms. Because lattice reorganization occurs on sub-picosecond timescales \textsuperscript{18-23}, most probes average over the very dynamics that define the quasiparticle.

Resolving polaron formation therefore requires a femtosecond probe that connects lattice fluctuations directly to electronic lineshapes. Coherent multidimensional spectroscopy (CMDS) provides access to homogeneous frequency–frequency correlations through the anti-diagonal (AD) linewidth of two-dimensional spectra\textsuperscript{24-26}. Measurements on perovskite quantum dots revealed narrow initial homogeneous widths that broaden rapidly with population time \textsuperscript{27-29}, while CdSe exhibits weak oscillatory modulation \textsuperscript{27,30} characteristic of underdamped phonons. These contrasting behaviors suggest fundamentally different lattice responses, but a microscopic framework linking the observed AD dynamics to Landau’s polaron picture \textsuperscript{1,2} has been lacking.

Here we establish that connection. Within lineshape theory, \textsuperscript{24-26,31} the population-time evolution of the anti-diagonal linewidth is governed directly by the decay of lattice frequency–frequency correlations. Underdamped phonon modes generate oscillatory modulation of the homogeneous linewidth, reflecting coherent lattice motion. In contrast, overdamped collective fluctuations produce monotonic growth governed by an exponential correlation decay growth arising from finite-time correlation decay. The latter behavior corresponds to the dynamical formation of a Landau polaron. The resulting correlation times and scaling trends agree quantitatively with multidimensional spectroscopic measurements, identifying anti-diagonal linewidth dynamics as a direct measure of polaron formation on the timescale of atomic motion.

Figure 1 delineates the two dynamical regimes central to this work: coupling to discrete lattice normal modes and coupling to a collectively reorganizing polarization field. In crystalline semiconductors such as CdSe, lattice excitations are well-defined harmonic phonons. The time-domain Raman spectrum, obtained by Fourier transforming transient absorption oscillations, displays sharp acoustic and optical modes with minimal low-frequency spectral weight, reflecting long vibrational coherence times and weak anharmonic coupling\textsuperscript{32}. In this regime, exciton–phonon interaction is well described by a displaced harmonic oscillator: photoexcitation impulsively shifts a normal-mode coordinate, generating a coherent vibrational wavepacket that oscillates reversibly about a new equilibrium. The transition frequency therefore retains phase memory on the phonon timescale, producing oscillatory modulation of the homogeneous linewidth that reflects structured lattice correlations\textsuperscript{27,30}.

Halide perovskites exhibit qualitatively different dynamics\textsuperscript{33}. Their vibrational spectral density is dominated by broad low-frequency weight extending to zero frequency, indicative of overdamped, diffusive lattice fluctuations rather than coherent normal modes. The relevant coordinate is not a single phonon but a collective polarization field that reorganizes following excitation\textsuperscript{13,27-29,34,35}. Instead of oscillatory motion, the lattice evolves toward a self-consistent distortion, and the transition frequency undergoes overdamped relaxational reorganization with exponentially decaying correlations. This overdamped regime is the dynamical setting in which a Landau polaron can form.

A number of ultrafast probes have been invoked as evidence of polaron formation\textsuperscript{6,13,20,21,34-45}. Yet for two fundamental reasons: their observables do not project onto the relevant order parameter, and their temporal resolution is insufficient to resolve the correlation dynamics.

Transient absorption (TA) measures population changes and differential optical density\textsuperscript{4,13,34-36,40-42}; coherent phonons observed in TA, particularly at low temperature, reflect underdamped normal-mode oscillations rather than the overdamped buildup of a collective polarization field. Ultrafast terahertz (THz) spectroscopy measures transient photoconductivity and effective mass renormalization\textsuperscript{44-46}, revealing carrier mobility and dielectric screening consistent with polaronic dressing, but it accesses transport properties rather than the decay of microscopic frequency correlations.

Angle-resolved photoemission spectroscopy (ARPES) detects band renormalization and effective mass changes\textsuperscript{5}, and ultrafast electron or X-ray diffraction (UED) measures lattice displacement and diffuse scattering\textsuperscript{20,21,38,39,43}.

While these observables could in principle be related to a polaron order parameter, no framework has connected them to the decay of correlation functions that govern quasiparticle formation. Moreover, typical instrument response functions exceeding ~100 fs are insufficient to resolve the ~150 fs correlation time characteristic of polaron formation. Quantitative observation further requires near-band-edge excitation at low fluence to avoid hot-carrier and multiparticle effects that obscure intrinsic lattice reorganization. As a result, the decay of frequency memory that defines polaron birth has remained experimentally inaccessible until recently\textsuperscript{27-29}.

Coherent multidimensional spectroscopy uniquely resolves the anti-diagonal (AD) linewidth, which directly reflects frequency–frequency correlations during the population time\textsuperscript{24-26,47}. In perovskite quantum dots, the AD linewidth evolves from an initially narrow homogeneous width to exponential broadening with systematic halide and size dependence\textsuperscript{27-29}, whereas CdSe exhibits weak oscillatory modulation\textsuperscript{27,30}. These contrasting behaviors point to distinct lattice responses. What has been lacking is a microscopic framework connecting this observable to Landau’s description of polaron formation. Establishing that connection is the objective of the theory developed below.

We now develop the microscopic framework linking lattice dynamics to the measured anti-diagonal (AD) linewidth. Within the Kubo–Anderson formalism, the population-time evolution of the AD linewidth is governed directly by the frequency–frequency correlation function \(C_{\omega}(t)\) of the transition energy\textsuperscript{24-26,31}. Underdamped phonon environments produce oscillatory, sign-changing correlations that manifest as reversible modulation of the homogeneous linewidth \textsuperscript{27,30}. In contrast, overdamped collective fluctuations lead to monotonic exponential decay of frequency correlations and corresponding monotonic linewidth growth\textsuperscript{27-29}. It is this overdamped correlation decay that characterizes the dynamical formation of a Landau polaron.

Polaron formation is the time-dependent emergence of a collective lattice polarization surrounding an electronic excitation\textsuperscript{1,2,5}. An exciton generates a local electric field \(E_{exc}(\mathbf{r})\) that couples to the lattice polarization density \(P(\mathbf{r},t)\), shifting the transition energy according to
\begin{equation}
\delta E(t) = - \int d^{3}r\, P(\mathbf{r},t) \cdot E_{exc}(\mathbf{r}).
\tag{1}
\end{equation}

Because the distortion involves many lattice degrees of freedom, we introduce a coarse-grained scalar order parameter \(\eta(t)\), defined as the polarization projected along the excitonic field and averaged over the polaron volume \(V_{pol}\),
\begin{equation}
\eta(t) = \frac{1}{V_{pol}}\int_{V_{pol}} d^{3}r\, \widehat{e} \cdot P(\mathbf{r},t).
\tag{2}
\end{equation}

This coordinate is not a normal phonon mode and, at room temperature, is generically overdamped. Substituting Eq. (2) into Eq. (1) yields a linear coupling between the order parameter and the transition frequency,
\begin{equation}
\delta\omega(t) = g\,\eta(t),
\tag{3}
\end{equation}
where \(g\) is an effective exciton–lattice coupling constant. The experimentally relevant quantity is the correlation function of these fluctuations,
\begin{equation}
C_{\omega}(t) = \langle\delta\omega(t)\delta\omega(0)\rangle = g^{2}\langle\delta\eta(t)\delta\eta(0)\rangle.
\tag{4}
\end{equation}

Linearizing fluctuations about the mean polarization yields relaxational dynamics of a coarse-grained order parameter. Near equilibrium, such overdamped Langevin dynamics reduce generically to a linear stochastic equation whose stationary solution is the Ornstein–Uhlenbeck process \textsuperscript{48},
\begin{equation}
\frac{d\,\delta\eta(t)}{dt} = - \frac{1}{\tau_{pol}}\delta\eta(t) + \xi(t),
\tag{5}
\end{equation}
with $\langle \xi(t)\rangle = 0$ and $\langle \xi(t)\xi(0)\rangle \propto \delta(t)$, corresponding to Gaussian white-noise driving of the overdamped order parameter. The resulting correlation function is
\begin{equation}
\langle\delta\eta(t)\delta\eta(0)\rangle = \langle\delta\eta^{2}\rangle e^{- |t|/\tau_{pol}}.
\tag{6}
\end{equation}

Within second-order cumulant theory\textsuperscript{31}, the optical coherence decay is determined by
\begin{equation}
g(t) = \int_{0}^{t} d\tau \int_{0}^{\tau} d\tau^{'}\, C_{\omega}(\tau^{'}) ,
\tag{7}
\end{equation}
and in two-dimensional spectroscopy the anti-diagonal linewidth isolates this homogeneous contribution. For exponential correlations, the AD width evolves as
\begin{equation}
\Gamma_{AD}(t_{2}) \approx \Gamma_{\infty}\sqrt{1 - e^{- t_{2}/\tau_{p}}}.
\tag{8}
\end{equation}

Equation~(8) follows from the Ornstein--Uhlenbeck description of frequency fluctuations, in which the frequency variance evolves as $1 - e^{-t_{2}/\tau_{p}}$. Because the anti-diagonal linewidth is proportional to the square root of this variance, the linewidth itself follows the square-root dependence given above. This expression represents the minimal result for an Ornstein–Uhlenbeck frequency-fluctuation process in the experimentally relevant limit where static inhomogeneous broadening exceeds the initial homogeneous linewidth \textsuperscript{31,49-51}. More complete treatments show that the absolute anti-diagonal linewidth at \(t_{2}=0\) depends on the full set of homogeneous and inhomogeneous broadening parameters and may exhibit weak deviations from strictly monotonic behavior as the static limit is approached. However, the characteristic growth time extracted from the linewidth evolution remains equal to the correlation time \(\tau_{p}\), which is the primary quantity of interest here.

Equation (8) therefore establishes the central result of this work: the population-time evolution of the anti-diagonal linewidth directly measures the decay of the order-parameter correlation function associated with collective lattice polarization. Monotonic growth of the linewidth is the dynamical signature of overdamped relaxational polarization dynamics, whereas underdamped phonons produce primarily oscillatory modulation without cumulative spectral diffusion. Because the anti-diagonal linewidth isolates homogeneous frequency–frequency correlations, the observation of monotonic linewidth growth implies a finite-time decay of a collective polarization field. The multidimensional measurements reported in 2019\textsuperscript{27} exhibit precisely this behavior on a ~150 fs timescale, thereby satisfying the theoretical criterion for resolving Landau-polaron formation in real time. The population-time evolution of the anti-diagonal linewidth is thus naturally described by an Ornstein–Uhlenbeck frequency-fluctuation process, providing a minimal stochastic representation of order-parameter formation in a photoexcited polar quantum material.

Figure 2 illustrates the stochastic dynamics implied by the overdamped order-parameter model. For exponential correlation decay with characteristic time \(\tau_{pol}\), the transition frequency exhibits colored noise with finite memory. Fig2a shows representative growth of the lattice order parameter following excitation, highlighting the emergence of a well-defined reorganization time. Fig2b presents sample trajectories generated by the Ornstein–Uhlenbeck process, demonstrating bounded stochastic fluctuations rather than coherent oscillations. Fig2c displays the corresponding autocorrelation functions, which converge to the exponential form \(C(t)=e^{- t/\tau_{pol}}\). Within cumulant theory \textsuperscript{31}, this decay directly produces the monotonic evolution of the anti-diagonal linewidth predicted by Eq. (8).

Figure 3 connects these stochastic correlations to multidimensional spectroscopy. Fig3a shows the CMDS pulse sequence, in which frequency correlations accumulated during the population time \(t_{2}\) determine the anti-diagonal width of the two-dimensional spectrum. Fig3b presents calculated spectra at early and late \(t_{2}\): initially elongated along the diagonal when frequency memory is preserved, and progressively circular as memory decays. This lineshape evolution reflects spectral diffusion governed by \(C(t_{2})\). Fig3c compares correlation functions for perovskite and CdSe quantum dots. Perovskites exhibit monotonic exponential decay consistent with overdamped collective dynamics, whereas CdSe displays weak decay with oscillatory structure characteristic of coherent LO phonons. The corresponding linewidth dynamics, computed from Eq. (8) and compared with experiment, reproduce the exponential broadening observed in perovskites and the oscillatory modulation seen in CdSe.

The coherent multidimensional measurements therefore resolve the decay of frequency correlations in real time. In perovskite quantum dots, the observed exponential anti-diagonal broadening signifies overdamped collective polarization dynamics, while CdSe remains in the underdamped phonon regime. The linewidth evolution thus reflects the emergence of a lattice distortion surrounding the exciton and provides direct access to the real-time formation of a Landau polaron.

In conclusion, the formation of a Landau polaron—the dynamical dressing of an electronic excitation by a self-consistent lattice distortion—has long been inferred from steady-state renormalization effects but not resolved in time. Conventional probes reveal transport renormalization, dielectric screening, or structural displacement, yet they do not isolate the decay of frequency correlations that governs quasiparticle emergence. Here we establish that the waiting-time evolution of the anti-diagonal linewidth in coherent multidimensional spectroscopy directly reflects the correlation function of a collective lattice polarization order parameter. Within this framework, monotonic growth governed by an exponential correlation decay growth is the necessary dynamical consequence of overdamped polarization-field relaxation. Because the multidimensional measurements reported in 2019 exhibit precisely this exponential correlation decay on a ~150 fs timescale, the observed anti-diagonal linewidth evolution constitutes a direct real-time observation of Landau polaron formation.

Supplementary Information.
Details of the theory and fitting are provided in the SI.

Acknowledgements.
Financial support from NSERC and McGill University is acknowledged.

\nocite{*}
\bibliographystyle{apsrev4-2}
\bibliography{references}

\begin{figure*}[p]
\centering
\vspace{-10.0 cm}
\includegraphics[width=0.96\textwidth,height=0.70\textheight,keepaspectratio]{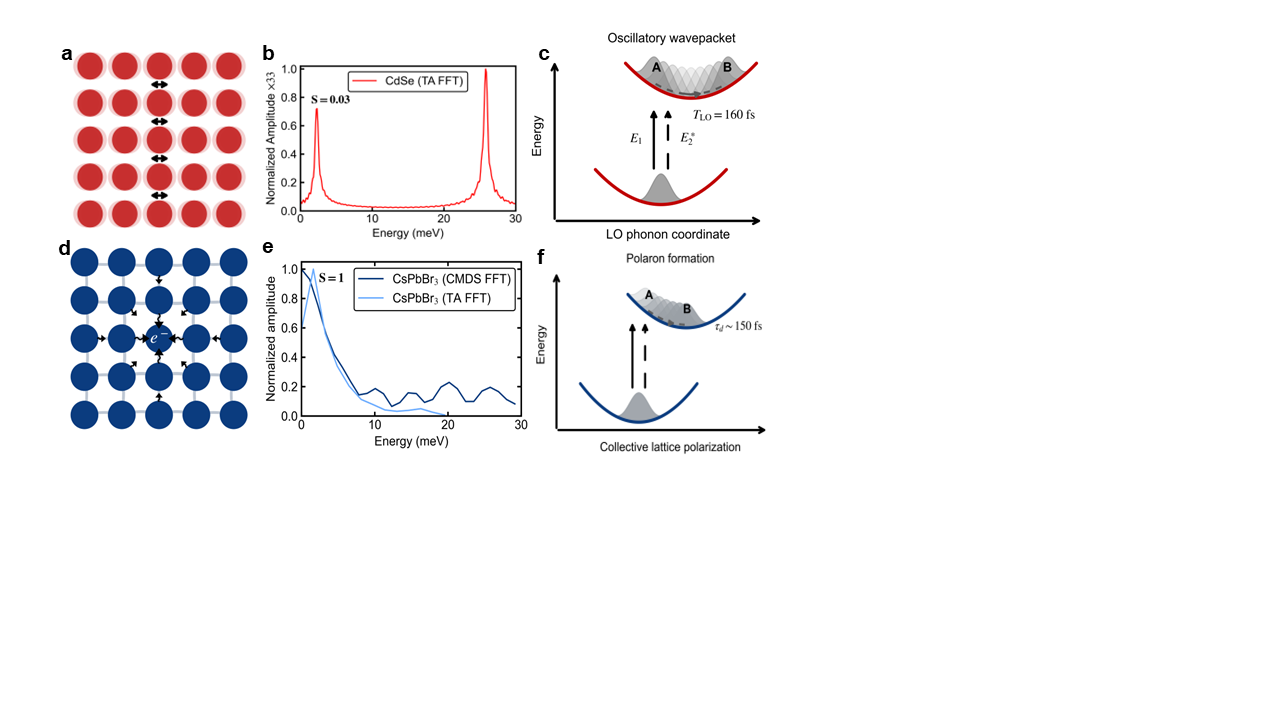}
\caption{Distinction between normal-mode phonons and collective polarization dynamics.
(a) In crystalline semiconductors such as CdSe, lattice excitations are well-defined normal modes. (b) Time-domain Raman spectrum of CdSe quantum dots obtained by Fourier transforming transient absorption oscillations, showing sharp acoustic and optical phonons with negligible low-frequency spectral weight. (c) Displaced harmonic oscillator model illustrating coherent phonon dressing: excitation impulsively shifts a normal-mode coordinate, producing reversible oscillatory modulation of the transition energy. (d) In halide perovskites, lattice response is dominated by dynamically soft, anharmonic fluctuations rather than discrete modes. (e) Corresponding spectral density extracted from CMDS and transient absorption exhibits broad low-frequency weight extending to zero frequency, indicative of overdamped fluctuations. (f) Configuration coordinate for a collective polarization field: following excitation, the lattice evolves diffusively toward a self-consistent distortion rather than oscillating coherently. This overdamped regime enables Landau polaron formation.}

\end{figure*}

\begin{figure}[p]
\centering
\includegraphics[width=\columnwidth,height=0.78\textheight,keepaspectratio]{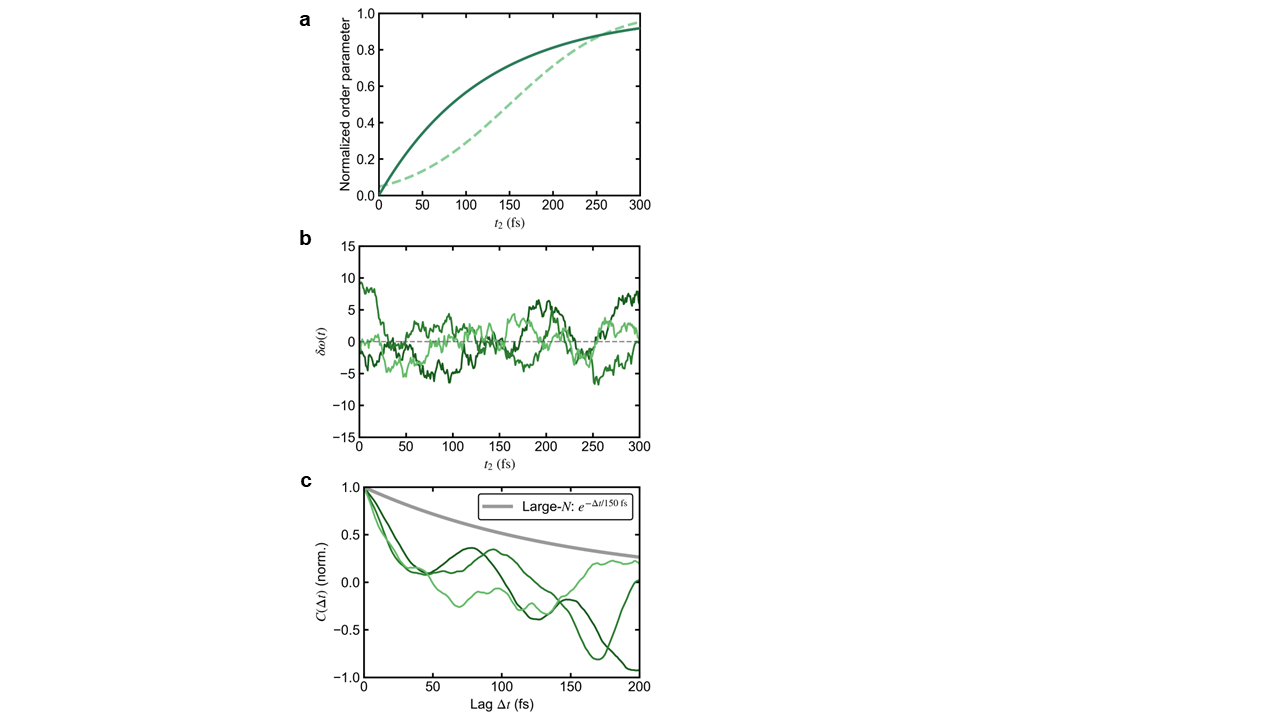}
\caption{ Stochastic dynamics of an overdamped lattice order parameter. (a) Representative time evolution of a normalized lattice order parameter following excitation, illustrating a characteristic reorganization time, following solution of Eq. (5). (b) Sample stochastic trajectories of transition-frequency fluctuations generated by an Ornstein–Uhlenbeck process with correlation time $\tau_{\mathrm{pol}}$. (c) Autocorrelation functions of the trajectories in (b), converging to exponential decay $C(t)=\exp(-t/\tau_{\mathrm{pol}})$. Exponential memory loss of the order parameter produces monotonic anti-diagonal linewidth growth within cumulant lineshape theory.}

\end{figure}

\begin{figure*}[p]
\centering
\includegraphics[width=0.96\textwidth,height=0.66\textheight,keepaspectratio]{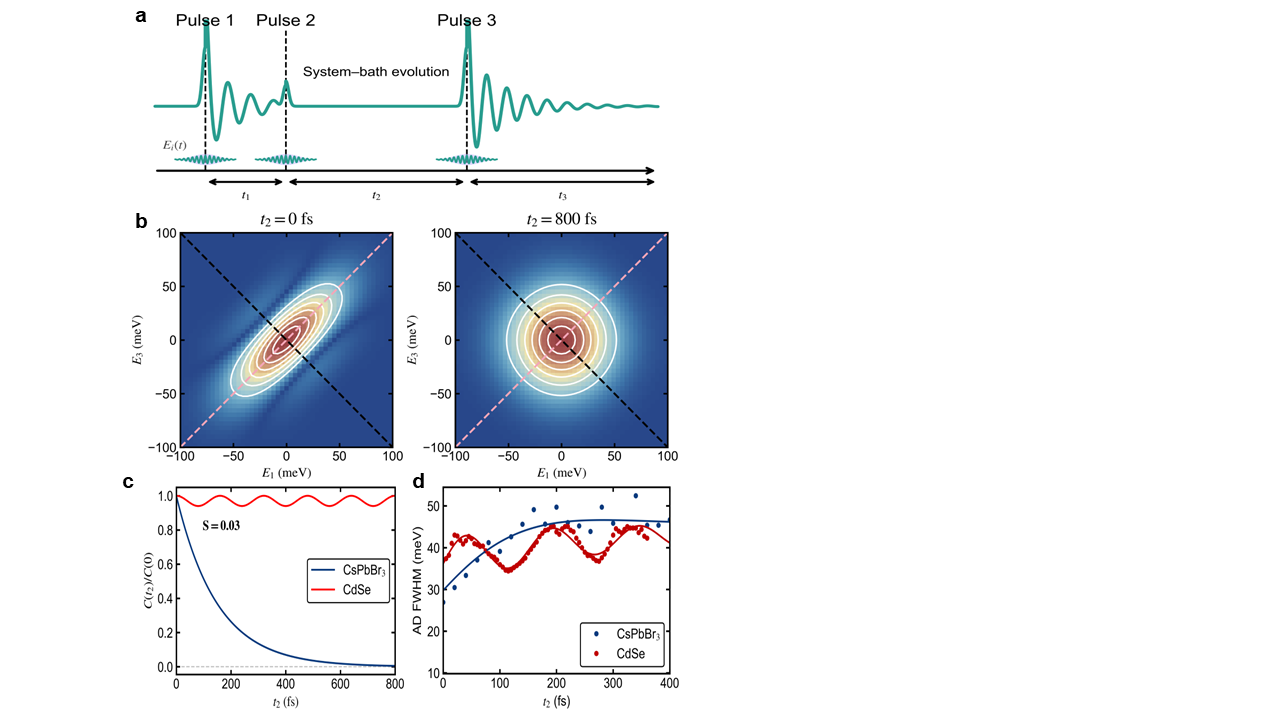}
\caption{Connection between correlation decay and multidimensional lineshape evolution. (a) CMDS pulse sequence consisting of three phase-resolved electric fields shown above the time axis. The first pulse creates a coherence, the second converts it to a population that evolves during waiting time $t_{2}$ under system–bath interactions, and the third converts the evolved state into a radiated third-order polarization. Frequency correlations accumulated during $t_{2}$ determine the anti-diagonal linewidth. (b) Calculated two-dimensional spectra at early and late $t_{2}$. Initially, excitation and emission frequencies are correlated, producing a diagonally elongated spectrum with narrow anti-diagonal width. As correlations decay, the spectrum becomes more circular and the anti-diagonal width increases, reflecting spectral diffusion governed by $C(t_{2})$. (c) Measured and modeled frequency–frequency correlation functions for perovskite and CdSe quantum dots. Perovskites exhibit monotonic exponential decay consistent with overdamped collective polarization dynamics, whereas CdSe shows weak decay with oscillatory structure characteristic of coherent LO phonons. (d) Population-time evolution of the anti-diagonal linewidth. The solid curves are calculated from Eq. (8) using the correlation functions in panel (c), and symbols represent experimental measurements. Perovskites exhibit monotonic linewidth growth governed by exponential correlation decay, whereas CdSe shows oscillatory modulation without cumulative broadening. The agreement demonstrates that the observed linewidth dynamics directly reflect the underlying correlation functions and constitute a real-time signature of Landau polaron formation.}

\end{figure*}


\end{document}


\textbf{Supporting Information}

\textbf{Real-Time Formation of a Landau Polaron}

Priya Nagpal\textsuperscript{1}, Arnab Ghosh\textsuperscript{1}, Hélène
Seiler\textsuperscript{2}, Samuel Palato\textsuperscript{3}, and
Patanjali Kambhampati\textsuperscript{1*}

\textsuperscript{1}Department of Chemistry, McGill University, Montreal,
H3A 0B8, Canada

\textsuperscript{2}Freie Universität Berlin, Arnimallee 14, 14195
Berlin, Germany

\textsuperscript{3}Department of Chemistry, Humbolt University, Berlin
12489, Germany

\href{mailto:*pat.kambhampati@mcgill.ca}{\nolinkurl{*pat.kambhampati@mcgill.ca}}

\textbf{\hfill\break}

{\small
\textbf{Table of contents.}

S1. Problem statement: polarons as quasiparticle \emph{birth} rather
than quasiparticle \emph{state}

S2. Microscopic coupling: polarization field, exciton field, and the
energetic meaning of ``polaron formation''

S3. Coarse-graining: why a polaron requires an order parameter and what
that order parameter actually is

S4. From field theory to an effective free energy for the order
parameter: where the Landau functional comes from

S5. Coarse-grained dynamics: why a Langevin equation is inevitable

S6. Linearization and the Ornstein--Uhlenbeck process: definition and
derivation

S7. From order-parameter correlations to frequency correlations and
spectral diffusion amplitude

S8. Raman spectrum, spectral density, and correlation functions: what is
constrained by experiment and what is not

S9. Cumulant expansion and lineshape theory: derivation of
\(e^{- g(t)}\)and the origin of \(g(t)\)

S10. Functional forms for \(g(t)\): delta-function, Kubo, underdamped
phonons, overdamped polarons

S11. Why 1D spectra cannot isolate order-parameter dynamics: mixing of
static disorder and spectral diffusion

S12. From response functions to 2D spectra: where ``diagonal'' and
``anti-diagonal'' are born mathematically

S13. Why \(\Gamma_{AD}(t_{2} = 0)\)is the homogeneous linewidth and why
spectral diffusion appears as time-dependent growth

S14. Contrast case: underdamped phonons and ``AC on DC'' oscillatory
linewidth modulation

S15. Time-resolution requirement: why the experiment must satisfy
\(IRF \ll \tau_{pol}\)

\textbf{\hfill\break}
}

\newpage

\textbf{S1. Problem statement: polarons as quasiparticle \emph{birth}
rather than quasiparticle \emph{state}}

A central habit in condensed matter theory is to treat quasiparticles as
stationary objects: defined by eigenstates, renormalized parameters, and
long-time response. In that tradition, the polaron is ``observed''
indirectly by static or quasi-static consequences, including effective
mass renormalization, mobility suppression, equilibrium spectral shifts,
or broadened absorption/emission. Those observables establish the
\emph{existence} of a polaronic dressing in the long-time limit, but
they do not identify a uniquely defined measurement principle for the
\textbf{formation event itself}, namely the time-dependent emergence of
the lattice polarization cloud following creation of a band-edge
excitation.

The goal of this SI is to present a microscopically anchored measurement
theory that answers a precise question: \textbf{what is the correct
order parameter for polaron formation, what is its correlation function,
and what experimental observable measures that correlation function in
real time?} The answer will be that time-resolved \textbf{anti-diagonal
linewidth dynamics} in \textbf{absorptive 2D electronic spectra} provide
direct access to the decay of frequency memory induced by the overdamped
order parameter. This transforms polaron formation from an inferred
state into a directly measurable dynamical process.

\textbf{\hfill\break}

\textbf{S2. Microscopic coupling: polarization field, exciton field, and
the energetic meaning of ``polaron formation''}

At the microscopic level the lattice response is described by a
polarization density field \(\mathbf{P}(\mathbf{r},t)\), which captures
the local dipole moment per unit volume arising from ionic displacements
and electronic screening. The electronic excitation (exciton) generates
a spatially localized electric field
\(\mathbf{E}_{exc}(\mathbf{r})\)determined by its charge distribution
and envelope function. The interaction energy between a polarization
field and an external electric field is standard electrodynamics. The
instantaneous energy shift of the electronic transition induced by
lattice polarization is therefore

\textbf{Microscopic exciton--polarization coupling}

\begin{equation}
\delta E(t) = - \int d^{3}r\,\mathbf{P}(\mathbf{r},t) \cdot \mathbf{E}_{exc}(\mathbf{r}).
\end{equation}

Equation (1) is the first conceptual hinge. It says: as the lattice
polarizes around the excitation, the transition energy is continuously
renormalized. In a ``static polaron'' view, one implicitly considers the
asymptotic value \(\delta E(\infty)\). In a ``birth'' view, the object
of interest is the full time-dependent stochastic process
\(\delta E(t)\)and, crucially, its fluctuations.

It is convenient to express the energy shift in frequency units,

\textbf{Frequency shift as spectroscopic noise}

\begin{equation}
\delta\omega(t) = \frac{\delta E(t)}{\hslash}.
\end{equation}

The central quantity that controls homogeneous linewidths in nonlinear
optical response is not \(\delta\omega(t)\)itself, but its two-time
correlation function \(\langle\delta\omega(t)\delta\omega(0)\rangle\).
The remainder of this SI is the rigorous route from microscopic lattice
physics to that correlation function and then to the directly measured
CMDS observable.

\textbf{\hfill\break}

\textbf{S3. Coarse-graining: why a polaron requires an order parameter
and what that order parameter actually is}

Tracking \(\mathbf{P}(\mathbf{r},t)\) exactly would require resolving a
high-dimensional phase space: many phonon branches, many wavevectors,
disorder, and anharmonic mode--mode couplings. This microscopic
description is neither experimentally accessible nor conceptually
faithful to a quasiparticle, because quasiparticles are low-dimensional
emergent variables describing collective behavior.

The correct language for ``collective birth'' is therefore \textbf{order
parameter dynamics}. The order parameter is a coarse-grained variable
that compresses microscopic details into a collective coordinate that is
slow, experimentally relevant, and causally linked to the measured
observable.

To define a meaningful coarse-graining, one must specify (i) a spatial
averaging scale, set here by the exciton extent and hence the polaron
volume, and (ii) a projection direction, set by the excitonic field /
transition dipole.

Let \(w(\mathbf{r})\)be a normalized weight function localized over the
polaron region, \(\int d^{3}r\text{ }w(\mathbf{r}) = 1\). Let
\(\widehat{\mathbf{e}}\)denote the direction of the relevant
polarization component (for an isotropic ensemble,
\(\widehat{\mathbf{e}}\)is effectively set by the optical transition
dipole and averaged). The coarse-grained order parameter is defined by
projecting \(\mathbf{P}\) onto \(\widehat{\mathbf{e}}\) and averaging over the polaron region:

\textbf{Polaron order parameter (coarse-grained polarization coordinate)}

\begin{equation}
\eta(t) \equiv \int d^{3}r\,w(\mathbf{r})\,\widehat{\mathbf{e}} \cdot \mathbf{P}(\mathbf{r},t).
\end{equation}

This definition is not arbitrary. It is a controlled projection in the
same sense as hydrodynamic coarse-graining: \(\eta(t)\)retains the
collective component of polarization that the exciton ``feels'', while
discarding microscopic detail irrelevant to spectroscopy. It is
important to state explicitly what \(\eta(t)\)is not: it is not a
normal-mode coordinate with a single frequency; it is not guaranteed to
oscillate; it is not tied to a harmonic potential. It is a collective
polarization coordinate in an abstract configuration space, analogous to
solvent polarization coordinates in solvation dynamics.

Now rewrite the microscopic coupling (1) in terms of the order
parameter. Decompose the excitonic field as
\(\mathbf{E}_{exc}(\mathbf{r}) = E_{0}f(\mathbf{r})\widehat{\mathbf{e}}\),
where \(f(\mathbf{r})\)captures spatial localization. Then

\begin{equation}
\delta E(t) = - E_{0}\int d^{3}r\,f(\mathbf{r})\,\widehat{\mathbf{e}} \cdot \mathbf{P}(\mathbf{r},t).
\end{equation}

If the weighting \(w(\mathbf{r})\) is chosen proportional to
\(f(\mathbf{r})\) (the natural choice that maximizes overlap with the
exciton), then the integral becomes proportional to \(\eta(t)\). One
obtains an effective linear coupling

\textbf{Reduction to linear exciton--order-parameter coupling}

\begin{equation}
\delta E(t) = - g\,\eta(t),
\end{equation}

or equivalently

\textbf{Frequency noise from order parameter}

\begin{equation}
\delta\omega(t) = g\,\eta(t),
\end{equation}

where \(g\)collects geometric overlap and screening. Equation (5) is
the bridge: polaron order parameter dynamics \emph{are}
transition-frequency dynamics.

Because spectroscopy is sensitive to fluctuations, decompose

\textbf{Mean plus fluctuation decomposition}

\begin{equation}
\eta(t) = \bar{\eta}(t) + \delta\eta(t),\qquad
\delta\omega(t) = g\,\delta\eta(t).
\end{equation}

The frequency--frequency correlation function is therefore

\textbf{Frequency correlation in terms of order-parameter correlation}

\begin{equation}
C_{\omega}(t) \equiv \langle\delta\omega(t)\delta\omega(0)\rangle =
g^{2}\langle\delta\eta(t)\delta\eta(0)\rangle \equiv g^{2}C_{\eta}(t).
\end{equation}

Up to this point, nothing has been assumed about functional forms. We
have only established a physically inevitable chain: microscopic
polarization → coarse-grained order parameter → frequency noise →
correlation function.

\textbf{\hfill\break}

\textbf{S4. From field theory to an effective free energy for the order
parameter: where the Landau functional comes from}

To proceed, we need dynamics. A standard and rigorous route is to begin
with a coarse-grained free-energy functional for the polarization field.
The minimal Landau--Ginzburg functional consistent with symmetry is

\textbf{Landau--Ginzburg free-energy functional for polarization field}

\begin{equation}
\mathcal{F[\mathbf{P}]} = \int d^{3}r\left[
\frac{a}{2} \mid \mathbf{P} \mid^{2} + \frac{b}{4} \mid \mathbf{P} \mid^{4}
+ \frac{\kappa}{2} \mid \nabla\mathbf{P} \mid^{2}
- \mathbf{E}_{exc}(\mathbf{r}) \cdot \mathbf{P}(\mathbf{r},t)
\right].
\end{equation}

The first three terms encode local stiffness, anharmonicity, and spatial
cost of polarization gradients. The last term is precisely the exciton
field coupling that drives polaron formation. This functional is the
continuum version of what microscopic lattice models imply once one
integrates over fast degrees of freedom and expands in long-wavelength
polarization fields. The key point is that polaron formation is
collective and long-wavelength compared to atomic scale, so such a
functional is the correct starting point for an order parameter
description.

Now project this functional onto the coarse-grained coordinate
\(\eta(t)\). Under the coarse-graining assumption that the relevant
response within the polaron volume is approximately uniform along
\(\widehat{\mathbf{e}}\), write
\(\mathbf{P}(\mathbf{r},t) \approx \widehat{\mathbf{e}}P_{\parallel}(t)\)in
the dominant region. Then \(\eta(t)\)is proportional to
\(P_{\parallel}(t)\), and the functional reduces to an effective 1D free
energy

\textbf{Effective free energy for the order parameter}

\begin{equation}
F(\eta) = \frac{k}{2}\eta^{2} + \frac{u}{4}\eta^{4} - h\,\eta.
\end{equation}

Here \(k\) and \(u\) are renormalized coefficients derived from
\(a,b,\kappa\) and the geometry of the polaron volume, while

\textbf{Generalized field from the exciton}

\begin{equation}
h \equiv \int d^{3}r\,w(\mathbf{r})\,\widehat{\mathbf{e}} \cdot \mathbf{E}_{exc}(\mathbf{r}).
\end{equation}

This is the exact mathematical answer to your complaint that ``\(5\)
came from nowhere'': it comes from projecting (9) onto the
exciton-selected coarse-grained coordinate.

The minimum of \(F(\eta)\)corresponds to the steady-state polarization
cloud. Polaron ``birth'' is the time-dependent approach to that minimum
starting from \(\eta(0) \approx 0\)immediately after excitation.

\textbf{\hfill\break}

\textbf{S5. Coarse-grained dynamics: why a Langevin equation is inevitable}

The next reduction is from field dynamics to an effective stochastic
equation of motion for \(\eta(t)\). The conceptual statement is simple:
\(\eta(t)\) is slow; everything else is fast. When one integrates out
fast degrees of freedom (the bath), one obtains friction and noise
acting on the slow coordinate. This is not an approximation unique to
this problem; it is the general structure of nonequilibrium statistical
physics (Mori--Zwanzig projection).

For a non-conserved order parameter, the simplest dynamics is
relaxational (Model A). The deterministic relaxation is downhill in free
energy, and the bath produces stochastic forcing. The resulting equation
is

\textbf{Model-A Langevin equation for order parameter}

\begin{equation}
\frac{d\eta}{dt} = - \Gamma\frac{\partial F(\eta)}{\partial\eta} + \xi(t).
\end{equation}

The noise \(\xi(t)\)has zero mean and, at thermal equilibrium, obeys
fluctuation--dissipation. For Markovian bath forcing this is

\textbf{Fluctuation--dissipation for white noise}

\begin{equation}
\langle\xi(t)\rangle = 0,\qquad
\langle\xi(t)\xi(0)\rangle = 2\Gamma k_{B}T\,\delta(t).
\end{equation}

This pair is the operational meaning of coarse-graining: all microscopic
complexity is hidden in a mobility \(\Gamma\)and a noise correlator fixed
by temperature.

Taking the derivative of (10) yields
\begin{equation}
\frac{\partial F}{\partial\eta} = k\eta + u\eta^{3} - h,
\end{equation}
so

\textbf{Explicit nonlinear Langevin dynamics}

\begin{equation}
\frac{d\eta}{dt} = - \Gamma(k\eta + u\eta^{3} - h) + \xi(t).
\end{equation}

Polaron formation corresponds to the deterministic drift toward the
stable minimum set by \(h\), while spectral diffusion and homogeneous
broadening originate from fluctuations about that drift.

\textbf{S6. Linearization and the Ornstein--Uhlenbeck process: definition and derivation}

Spectroscopic linewidths are controlled by fluctuations around the mean.
Let \(\overset{ˉ}{\eta}\)be the stable minimum of \(F(\eta)\),
determined by

\textbf{Stationary condition for the mean}

\begin{equation}
\left.\frac{\partial F}{\partial \eta}\right|_{\overline{\eta}} = 0
\;\Rightarrow\;
k\,\overline{\eta} + u\,\overline{\eta}^{3} - h = 0 .
\end{equation}

where $\eta(t) = \overline{\eta} + \delta\eta(t)$ and expand
$F(\eta)$ around $\overline{\eta}$. The harmonic approximation is

\begin{equation}
F(\eta) \approx F(\overline{\eta}) + \frac{k_{\mathrm{eff}}}{2}\,(\delta\eta)^2 .
\end{equation}

where the curvature is

\textbf{Effective stiffness}

\begin{equation}
k_{\mathrm{eff}}
=
\left.\frac{\partial^{2}F}{\partial \eta^{2}}\right|_{\bar{\eta}}
=
k + 3u\,\bar{\eta}^{2}
\end{equation}

Substituting into the Langevin equation gives the linear stochastic equation

\textbf{Ornstein--Uhlenbeck equation (definition)}

\begin{equation}
\frac{d\,\delta\eta}{dt} = - \gamma\,\delta\eta(t) + \xi(t),\qquad
\gamma \equiv \Gamma k_{eff}.
\end{equation}

Equation (17) with white noise defines the Ornstein--Uhlenbeck (OU)
process: it is the unique stationary Gaussian Markov process with
linear restoring force and delta-correlated noise.

Define the polaron correlation time

\textbf{Polaron correlation time}

\begin{equation}
\tau_{pol} \equiv \gamma^{- 1}.
\end{equation}

Now solve for \(t > 0\):

\begin{equation}
\delta\eta(t) = \delta\eta(0)e^{- \gamma t} + \int_{0}^{t}ds\,e^{- \gamma(t - s)}\xi(s).
\end{equation}

Multiplying by \(\delta\eta(0)\)and averaging gives

\begin{equation}
\langle\delta\eta(t)\delta\eta(0)\rangle = \langle\delta\eta(0)^{2}\rangle e^{- \gamma t}.
\end{equation}

Thus

\textbf{OU correlation function}

\begin{equation}
C_{\eta}(t) \equiv \langle\delta\eta(t)\delta\eta(0)\rangle
= \langle\delta\eta^{2}\rangle e^{- t/\tau_{pol}}.
\end{equation}

The variance is fixed by thermal equilibrium in the harmonic well:

\textbf{OU variance from equipartition}

\begin{equation}
\langle\delta\eta^{2}\rangle = \frac{k_{B}T}{k_{eff}}.
\end{equation}

Equations (21--22) are not assumed; they follow explicitly from the OU
definition plus equilibrium.

\textbf{\hfill\break}

\textbf{S7. From order-parameter correlations to frequency correlations and spectral diffusion amplitude}

Using \(\delta\omega(t) = g\,\delta\eta(t)\), the frequency
correlation function is

\textbf{Frequency--frequency correlation function (FFCF)}

\begin{equation}
C_{\omega}(t) \equiv \langle\delta\omega(t)\delta\omega(0)\rangle = g^{2}C_{\eta}(t).
\end{equation}

Substituting gives

\textbf{Polaron FFCF as exponential memory loss}

\begin{equation}
C_{\omega}(t) = \Delta^{2}e^{- t/\tau_{pol}},\qquad
\Delta^{2} \equiv g^{2}\langle\delta\eta^{2}\rangle.
\end{equation}

The two parameters \(\Delta\)and \(\tau_{pol}\)are the entire content of
overdamped spectral diffusion at this level: \(\Delta\)is the spectral
diffusion amplitude (frequency-noise strength) and \(\tau_{pol}\) is the
correlation time, i.e. the polaron birth timescale.

\textbf{\hfill\break}

\textbf{S8. Raman spectrum, spectral density, and correlation functions:
what is constrained by experiment and what is not}

Experiment provides a vibrational spectrum, but one must be careful
about which theoretical object it constrains. In system--bath language,
what enters lineshape theory is the \textbf{bath spectral density}
\(J(\omega)\), which weights how strongly bath modes at frequency
\(\omega\)couple to the transition frequency. The equilibrium frequency
correlation function is related to \(J(\omega)\)through standard
fluctuation--dissipation relations.

Define a bath coordinate \(Q\)coupled linearly to the transition
frequency,

\textbf{Linear system--bath coupling}

\begin{equation}
\delta\omega = \sum_{k}c_{k}q_{k} \equiv Q,
\end{equation}

where \(q_{k}\)are bath mode coordinates and \(c_{k}\)are coupling
constants. Then the spectral density is

\textbf{Spectral density}

\begin{equation}
J(\omega) = \frac{\pi}{2}\sum_{k}\frac{c_{k}^{2}}{m_{k}\omega_{k}}\,\delta(\omega - \omega_{k}).
\end{equation}

The corresponding correlation function can be written as

\textbf{Correlation function from spectral density (quantum form)}

\begin{equation}
C_{\omega}(t) = \frac{1}{\pi}\int_{0}^{\infty}d\omega\,J(\omega)\,
\coth\!\left( \frac{\hslash\omega}{2k_{B}T} \right)\cos(\omega t).
\end{equation}

At high temperature \(k_{B}T \gg \hslash\omega\)over the relevant
low-frequency band,
\(\coth(\hslash\omega/2k_{B}T) \approx 2k_{B}T/\hslash\omega\), yielding
the classical form

\textbf{High-temperature correlation from spectral density}

\begin{equation}
C_{\omega}(t) \approx \frac{2k_{B}T}{\pi\hslash}\int_{0}^{\infty}d\omega\,
\frac{J(\omega)}{\omega}\cos(\omega t).
\end{equation}

This is the formal expression that underlies the practical statement you use: Fourier-transforming the experimentally determined low-frequency spectral density yields the correlation function. The key subtlety is that a CW Raman spectrum is not automatically \(J(\omega)\); impulsive Raman (time-domain) can be more faithful at low frequency, and the mapping depends on selection rules and response functions. In the SI, the right way to phrase it is: experiment constrains the effective low-frequency spectral density relevant to frequency fluctuations, and that constraint fixes the timescale \(\tau_{\mathrm{pol}}\) and the presence/absence of low-frequency continuum.
 

\textbf{\hfill\break}

\textbf{S9. Cumulant expansion and lineshape theory: derivation of}
\(\mathbf{e}^{\mathbf{- g(t)}}\)\textbf{and the origin of}
\(\mathbf{g(t)}\)

The optical polarization depends on the phase accumulated by frequency
fluctuations during coherence evolution. Define the stochastic phase

\textbf{Phase functional of frequency noise}

\begin{equation}
\phi(t) \equiv \int_{0}^{t}d\tau\,\delta\omega(\tau).
\end{equation}

The coherence factor is
\[
\langle e^{- i\phi(t)}\rangle =
\left\langle\exp\!\left[ - i\int_{0}^{t}d\tau\,\delta\omega(\tau) \right]\right\rangle.
\]

If \(\delta\omega(t)\)is Gaussian (OU is Gaussian), then \(\phi(t)\)is
Gaussian with zero mean, and the characteristic function identity gives

\textbf{Exact Gaussian cumulant identity}

\begin{equation}
\langle e^{- i\phi(t)}\rangle =
\exp\!\left[ - \frac{1}{2}\langle\phi(t)^{2}\rangle \right].
\end{equation}

Compute \(\langle\phi(t)^{2}\rangle\):

\begin{equation}
\langle\phi(t)^{2}\rangle = \int_{0}^{t}d\tau\int_{0}^{t}d\tau'\,
\langle\delta\omega(\tau)\delta\omega(\tau')\rangle
= \int_{0}^{t}d\tau\int_{0}^{t}d\tau'\,C_{\omega}(|\tau-\tau'|).
\end{equation}

Rewrite this in the standard one-sided form:

\textbf{Lineshape function definition}

\begin{equation}
g(t) \equiv \int_{0}^{t} d\tau \int_{0}^{\tau} d\tau'\, C_{\omega}(\tau').
\end{equation}

Then

\textbf{Coherence decay in cumulant form}

\begin{equation}
\left\langle\exp\!\left[ - i\int_{0}^{t}d\tau\,\delta\omega(\tau) \right]\right\rangle
= e^{- g(t)}.
\end{equation}

This is the mathematical core of Kubo--Mukamel lineshape theory. It is not an ansatz; for Gaussian noise it is exact, and for weakly non-Gaussian noise it is the controlled second-cumulant approximation.

\textbf{\hfill\break}

\textbf{S10. Functional forms for} \(\mathbf{g(t)}\)\textbf{:
delta-function, Kubo, underdamped phonons, overdamped polarons}

The power of the cumulant form is that different microscopic physics is encoded
entirely through \(C_{\omega}(t)\).

If \(C_{\omega}(t) = 2\Gamma_{\phi}\delta(t)\), then \(g(t)=\Gamma_{\phi}t\).

If \(C_{\omega}(t) = \Delta^{2}e^{- t/\tau_{c}}\), then inserting gives

\textbf{Kubo lineshape function}

\begin{equation}
g(t) = \Delta^{2}\tau_{c}^{2}\left( e^{- t/\tau_{c}} + \frac{t}{\tau_{c}} - 1 \right).
\end{equation}

For underdamped phonons, the correlation is oscillatory,

\textbf{Underdamped phonon FFCF}

\begin{equation}
C_{\omega}(t) = \Delta^{2}\cos(\Omega t)\,e^{- t/T_{2}}.
\end{equation}

which yields an oscillatory $g(t)$ and therefore oscillatory modulation of homogeneous response, rather than monotonic broadening.

For overdamped polarons, the OU form (25) is identical to Kubo exponential memory loss. Thus the ``polaron birth'' picture is precisely the statement that at room temperature the relevant coarse-grained polarization coordinate is overdamped and therefore has exponential memory loss, producing monotonic linewidth growth in the waiting time.

\textbf{\hfill\break}

\textbf{S11. Why 1D spectra cannot isolate order-parameter dynamics:
mixing of static disorder and spectral diffusion}

A 1D absorption or pump--probe spectrum collapses the response onto a
single frequency axis. In an ensemble, the transition frequency can be
decomposed as

\textbf{Static + dynamic decomposition}

\begin{equation}
\omega(t) = \omega_{0} + \delta\omega_{stat} + \delta\omega_{dyn}(t).
\end{equation}
Static inhomogeneity elongates a distribution of $\omega$ across the ensemble and contributes to apparent linewidth. Dynamic noise contributes to homogeneous broadening and spectral diffusion. In 1D, these effects are convolved. Consequently, even if polaron formation drives a dramatic time-dependent homogeneous broadening, 1D projections tend to express it as ambiguous changes in peak widths, shifts, or apparent kinetics mixed with cooling, trapping, and population evolution.

A 2D spectrum is the minimal object that cleanly separates static and dynamic contributions geometrically through correlated excitation and emission frequencies. That is why 2D is not ``extra information''; it is the correct information for the order-parameter problem.

\textbf{\hfill\break}

\textbf{S12. From response functions to 2D spectra: where ``diagonal''
and ``anti-diagonal'' are born mathematically}

In third-order nonlinear spectroscopy, the signal is generated by the third-order polarization 
$P^{(3)}(t_{3},t_{2},t_{1})$, where $t_{1}$ and $t_{3}$ are coherence periods and $t_{2}$ is the waiting time. 
The absorptive 2D spectrum is formed by Fourier transforming the signal with respect to $t_{1}$ and $t_{3}$ 
and taking the appropriate combination of rephasing and nonrephasing contributions:

\textbf{Absorptive 2D spectrum definition}

\begin{equation}
S_{abs}(\omega_{3},t_{2},\omega_{1}) =
\mathfrak{R}\left[
\int_{0}^{\infty}dt_{3}\int_{0}^{\infty}dt_{1}\,
e^{+ i\omega_{3}t_{3}}e^{- i\omega_{1}t_{1}}
\left( P_{R}^{(3)}+P_{NR}^{(3)} \right)
\right].
\end{equation}

The physical point is that the 2D spectrum is a joint distribution over excitation and emission frequencies, $(\omega_{1},\omega_{3})$. Its geometry encodes how much of the initial frequency is remembered after time $t_{2}$. The normalized memory factor is

\textbf{Frequency memory factor}

\begin{equation}
M(t_{2}) = \frac{C_{\omega}(t_{2})}{C_{\omega}(0)}.
\end{equation}

When $M(t_{2})=1$, excitation and emission are perfectly correlated and the peak is elongated along the diagonal. When $M(t_{2})=0$, correlation is lost and the peak becomes more anti-diagonally broadened.  
A transparent and standard representation is to model the joint distribution of $\delta\omega_{1}$ and $\delta\omega_{3}$ as a correlated Gaussian with covariance

\textbf{Two-time covariance matrix}

\begin{equation}
\Sigma(t_{2}) =
\begin{pmatrix}
\sigma^{2} & \sigma^{2}M(t_{2}) \\
\sigma^{2}M(t_{2}) & \sigma^{2}
\end{pmatrix},\qquad
\sigma^{2} \equiv C_{\omega}(0).
\end{equation}

Now define diagonal and anti-diagonal coordinates

\textbf{Coordinate transform}

\begin{equation}
\delta\omega_{D} = \frac{\delta\omega_{3} + \delta\omega_{1}}{\sqrt{2}},\qquad
\delta\omega_{AD} = \frac{\delta\omega_{3} - \delta\omega_{1}}{\sqrt{2}}.
\end{equation}

Compute their variances using (41). The results are

\textbf{Diagonal and anti-diagonal variances}

\begin{equation}
Var(\delta\omega_{D}) = \sigma^{2}(1 + M(t_{2})),\qquad
Var(\delta\omega_{AD}) = \sigma^{2}(1 - M(t_{2})).
\end{equation}

This derivation is the rigorous geometric origin of diagonal and anti-diagonal broadening in 2D.
Homogeneous broadening adds an additional baseline variance $\sigma_{\mathrm{hom}}^{2}$ that appears in all directions. Thus the total anti-diagonal variance can be written as

\textbf{Anti-diagonal variance including homogeneous baseline}

\begin{equation}
\sigma_{AD}^{2}(t_{2}) = \sigma_{\hom}^{2} + \sigma^{2}\left( 1 - M(t_{2}) \right).
\end{equation}

Expressed as FWHM linewidths (with the standard Gaussian conversion factor), this yields a directly testable prediction for $\Gamma_{AD}(t_{2})$.
For OU/polaron dynamics, $M(t_{2}) = e^{-t_{2}/\tau_{\mathrm{pol}}}$, giving

\textbf{Polaron prediction for anti-diagonal linewidth growth}

\begin{equation}
\sigma_{AD}^{2}(t_{2}) = \sigma_{\hom}^{2} + \sigma^{2}\left( 1-e^{- t_{2}/\tau_{pol}} \right).
\end{equation}

Equation (45) is the final operational mapping: it connects the polaron order-parameter correlation time $\tau_{\mathrm{pol}}$ and amplitude $\sigma^{2}$ to a directly measured, time-resolved anti-diagonal linewidth.

\textbf{\hfill\break}

\textbf{S13. Why}
\(\mathbf{\Gamma}_{\mathbf{AD}}\mathbf{(}\mathbf{t}_{\mathbf{2}}\mathbf{= 0)}\)\textbf{is
the homogeneous linewidth and why spectral diffusion appears as
time-dependent growth}

At $t_{2}=0$, the system has not yet undergone waiting-time evolution that destroys frequency memory. In this limit, the anti-diagonal direction isolates homogeneous broadening because static inhomogeneity is aligned along the diagonal. Mathematically, (S37) gives
$ \mathrm{Var}(\delta\omega_{AD}) = \sigma^{2}(1 - M) \rightarrow 0$ when $M \rightarrow 1$, leaving only the homogeneous baseline $\sigma_{\mathrm{hom}}^{2}$. Thus

\textbf{Anti-diagonal width at zero waiting time}

\begin{equation}
\sigma_{AD}^{2}(0) = \sigma_{\hom}^{2}.
\end{equation}

As $t_{2}$ increases, memory decays, $M(t_{2}) \rightarrow 0$, and the anti-diagonal variance approaches

\textbf{Long-time anti-diagonal width}

\begin{equation}
\sigma_{AD}^{2}(\infty) = \sigma_{\hom}^{2} + \sigma^{2}.
\end{equation}
Thus the growth of the anti-diagonal linewidth from $t_{2}=0$ to long $t_{2}$ is a direct measurement of the dynamic spectral diffusion amplitude $\sigma^{2}=C_{\omega}(0)$, and the time constant of that growth is $\tau_{\mathrm{pol}}$.

This is the precise theoretical meaning of your experimental observation: polaron formation appears as time-dependent growth of the homogeneous linewidth as the order parameter evolves and loses memory.

\textbf{\hfill\break}

\textbf{S14. Contrast case: underdamped phonons and ``AC on DC''
oscillatory linewidth modulation}

For CdSe-like systems with well-defined phonon modes, the bath is underdamped and correlations are oscillatory rather than purely relaxational. A minimal representation is

\begin{equation}
C_{\omega}(t) = \sum_{j}\Delta_{j}^{2}\cos(\Omega_{j}t)\,e^{- t/T_{2,j}}.
\end{equation}
Then the memory factor $M(t_{2})$ acquires oscillatory components, and (44) predicts oscillatory anti-diagonal linewidth modulation rather than monotonic growth. This explains why coherent phonons yield AC oscillations about a constant baseline, and why the presence of coherent phonons in 1D pump--probe does not automatically imply the same observable content as a 2D anti-diagonal linewidth measurement: the 1D projection mixes static and dynamic effects, whereas the 2D geometry isolates memory decay.

In the polaron case, the low-frequency continuum produces overdamped OU-like behavior, so $M(t_{2})$ decays exponentially and the anti-diagonal linewidth grows monotonically.

\textbf{\hfill\break}

\textbf{S15. Time-resolution requirement: why the experiment must
satisfy} \(\mathbf{IRF}\mathbf{\ll}\mathbf{\tau}_{\mathbf{pol}}\)

The measured $\Gamma_{AD}(t_{2})$ is the convolution of the true waiting-time dependence with the instrument response function (IRF). If the underlying dynamics has characteristic time $\tau_{pol}$, then resolving it requires the IRF to be significantly shorter. A practical criterion for simply distinguishing a rise is $IRF \lesssim \tau_{pol}/3$. A stricter criterion for faithful reconstruction of functional form is $IRF \lesssim \tau_{pol}/10$.

For room-temperature perovskite polaron formation with $\tau_{pol} \sim 150\,\mathrm{fs}$ as constrained by low-frequency spectral density, the strict criterion implies an IRF on the order of $\sim 15\,\mathrm{fs}$. This is precisely the experimental regime where CMDS operates and where many other ultrafast methods remain marginal or insufficient in practice, especially when combined with additional complications of non-band-edge excitation, high-fluence plasma regimes, and low-temperature measurements where the relevant overdamped polaron physics is suppressed.

This time-resolution criterion is not rhetorical; it is a direct statement about convolution and identifiability of $\tau_{pol}$ from measured dynamics.